\documentclass[conference]{IEEEtran}
\pdfoutput=1
\usepackage[]{hyperref}
\IEEEoverridecommandlockouts{}
\usepackage{cite}
\usepackage{amsmath,amssymb,amsfonts}
\usepackage{trimclip}
\usepackage{algorithm}
\usepackage{algorithmic}
\usepackage{graphicx}
\usepackage{textcomp}
\usepackage{xcolor}
\usepackage[a4paper, total={184mm,239mm}]{geometry}
\def\BibTeX{{\rm B\kern-.05em{\sc i\kern-.025em b}\kern-.08em
    T\kern-.1667em\lower.7ex\hbox{E}\kern-.125emX}}
\usepackage{paralist}
\usepackage{xspace}
\usepackage{makecell}
\usepackage{pifont}

\usepackage{booktabs}
\usepackage{enumitem}
\usepackage{microtype}
\usepackage{wasysym}

\usepackage[binary-units]{siunitx}

\usepackage{pgfplots}
\pgfplotsset{compat=1.17}
\usetikzlibrary{shapes,snakes}
\usetikzlibrary{angles,quotes}
\usetikzlibrary{calc}
\usetikzlibrary{positioning}
\pgfmathsetseed{2}
\newcommand\set[8]{% radius, irregularity
  \coordinate (center) at (#1,#2);
  \draw[#6,smooth cycle,tension=6,rounded corners=0.5mm] (center)
  \pgfextra {\pgfmathsetmacro\len{(#4)}}
  +(0:\len pt)
  \foreach \a in {10,20,...,80}{
    \pgfextra {\pgfmathsetmacro\len{{(#3) * (sin(\a) * sin(\a)) + (#4) * (cos(\a) * cos(\a)) +rand*(#5)}}}
    -- + (\a:\len pt)
  }
  \pgfextra {\pgfmathsetmacro\len{(#3)}}
  -- +(90:\len pt)
  \foreach \a in {100,110,...,170}{
    \pgfextra {\pgfmathsetmacro\len{{(#3) * (sin(\a) * sin(\a)) + (#4) * (cos(\a) * cos(\a)) +rand*(#5)}}}
    -- +(\a:\len pt)
  }
  \pgfextra {\pgfmathsetmacro\len{(#4)}}
  -- +(180:\len pt)
  \foreach \a in {190,200,...,260}{
    \pgfextra {\pgfmathsetmacro\len{{(#3) * (sin(\a) * sin(\a)) + (#4) * (cos(\a) * cos(\a)) +rand*(#5)}}}
    -- +(\a:\len pt)
  }
  \pgfextra {\pgfmathsetmacro\len{(#3)}}
  -- +(260:\len pt)
  \foreach \a in {280,290,...,350}{
    \pgfextra {\pgfmathsetmacro\len{{(#3) * (sin(\a) * sin(\a)) + (#4) * (cos(\a) * cos(\a)) +rand*(#5)}}}
    -- +(\a:\len pt)
  }
  -- cycle;
  \node[ellipse, minimum width=#4 * 2, minimum height =#3 * 2, color=#6] (#7) at (center) {#8};
}

\tikzset{>=latex}

\newcommand{\ie}{{\it i.e.}\xspace}
\newcommand{\eg}{{\it e.g.}\xspace}

\begin{document}

\title{Synthesizing Hardware-Software Leakage Contracts for RISC-V Open-Source Processors
\thanks{This project has received funding from the {European Research Council} under the European Union's Horizon 2020 research and innovation programme (grant agreement No. {101020415}), from the Spanish Ministry of Science, Innovation, and University under the projects PID2022-142290OB-I00 ESPADA and TED2021-132464B-I00 PRODIGY and  under the Ram\'on y Cajal grant RYC2021-032614-I, and from a gift by Intel Corporation.}
}
\author{\IEEEauthorblockN{Gideon Mohr}
\IEEEauthorblockA{
\textit{Saarland University}\\\textit{Saarland Informatics Campus}\\
Saarbrücken, Germany}
\vspace{-3mm}
\and
\IEEEauthorblockN{Marco Guarnieri}
\IEEEauthorblockA{
\textit{IMDEA Software Institute}\\
Madrid, Spain}
\and
\IEEEauthorblockN{Jan Reineke}
\IEEEauthorblockA{
\textit{Saarland University}\\\textit{Saarland Informatics Campus}\\
Saarbrücken, Germany}
\vspace{-3mm}
}

\maketitle

\begin{abstract}
    Microarchitectural attacks compromise security by exploiting software-visible artifacts of microarchitectural optimizations such as caches and speculative execution.
    Defending against such attacks at the software level requires an appropriate abstraction at the instruction set architecture (ISA) level that captures microarchitectural leakage.
    Hardware-software leakage contracts have recently been proposed as such an abstraction.
    
    In this paper, we propose a semi-automatic methodology for synthesizing hardware-software leakage contracts for open-source microarchitectures.
    For a given ISA, our approach relies on human experts to (a) capture the space of possible contracts in the form of contract templates and (b)  devise a test-case generation strategy to explore a microarchitecture's potential leakage.
    For a given implementation of an ISA, these two ingredients are then used to automatically synthesize the most precise leakage contract that is satisfied by the microarchitecture.

    We have instantiated this methodology for the RISC-V ISA and applied it to the Ibex and CVA6 open-source processors.
    Our experiments demonstrate the practical applicability of the methodology and uncover subtle and unexpected leaks.
\end{abstract}

\section{Introduction}\label{sec:intro}

Microarchitectural attacks~\cite{spectre2019,meltdown2018,Bulck2018,RIDL,Yarom14} compromise the security of programs by exploiting software-visible artifacts of microarchitectural optimizations such as caches and speculative execution.
Defending against such attacks in software is challenging:  instruction set architectures (ISAs), the traditional hardware-software interface, abstract from microarchitectural details and thus do not give any guarantees w.r.t.\ these attacks.

Hardware-software leakage contracts~\cite{contracts2021,wang2023specification} (short: leakage contracts) have recently been proposed as a new security abstraction at the ISA level to fill this gap. 
Such contracts aim to capture possible microarchitectural side-channel leaks by associating leakage traces, i.e., sequences of leakage observations, with ISA-level executions.
For example, a contract could expose the addresses of memory instructions as leakage observations to capture data cache leaks. 
Similarly, a contract could expose the operands of division instructions to capture leakage via operand-dependent latencies. 
Given a contract that faithfully captures the microarchitectural leakage of a processor, it is then possible to program the hardware securely by making sure that all leakage observations are independent of secrets~\cite{contracts2021}.

However, most of today's processor designs lack precise specifications of  microarchitectural leakage. 
In this work, we close this gap by proposing a semi-automatic methodology for synthesizing leakage contracts from open-source processor designs.\looseness=-1

Our methodology consists of four steps:\looseness=-1 
\begin{asparaenum}
    \item {\bf Definition of Contract Template}. 
        A human expert determines a set of \emph{contract atoms} that capture potential instruction-level leakage observations. 
        For example, a contract atom may expose the value of the register operand of a memory instruction.
        The set of all contract atoms forms the \emph{contract template}, and any of its subsets is a candidate \emph{contract}.
    \item {\bf Test-Case Generation}.
        A human expert devises a test-case generation strategy.
        Each \emph{test case} consists of two ISA-level programs with fixed data inputs. 
        These test cases are used to exercise and analyze a processor's microarchitectural leakage. 
    \item {\bf Evaluation of Test Cases}.
        Test cases are automatically evaluated on  the target processor design  to determine which test cases are \emph{distinguishable}, \ie, lead to distinguishable executions for a given microarchitectural attacker. Distinguishable test cases expose actual leaks that need to be accounted for at contract level.
        Test cases are also automatically evaluated on the contract template to derive the set of \emph{distinguishing atoms}, i.e., those atoms that distinguish the two programs.
        For instance, an atom exposing the value of register operands for memory instructions will distinguish programs accessing different addresses.

    \item {\bf Automatic Contract Synthesis}.
        Based on the results of the test-case evaluation, a {contract}, \ie a set of contract atoms, is automatically synthesized from the distinguishing atoms associated with attacker-distinguishable test case.
        Our approach ensures that the synthesized contract is \emph{satisfied} by the processor design on all test cases, \ie, it captures all leaks exposed by the test cases on the processor.
        Moreover, it also ensures that the synthesized contract is the \emph{most precise} such contract, \ie, it distinguishes the fewest attacker-indistinguishable test cases.
\end{asparaenum}

\smallskip

The proliferation of the open-source RISC-V ISA~\cite{Asanovic14} and its growing ecosystem is promising in this context for two reasons:
\begin{inparaenum}[(1)]
    \item The simplicity and modularity of the ISA provides a good foundation for the definition of contract atoms.
    \item The growing body of open-source RISC-V processor designs of varying complexity provides natural targets for contract synthesis.
\end{inparaenum}
Leveraging the RISC-V Formal Interface~\cite{wolf2017riscv-formal}, we have instantiated our contract-synthesis methodology for the RISC-V ISA and applied it to two open-source processor designs: Ibex~\cite{ibex} and CVA6~\cite{cva6}.
Our experiments reveal subtle previously undocumented cases of microarchitectural leakage and demonstrate the practical applicability of our methodology.\looseness=-1
\enlargethispage{0.4\baselineskip}

To summarize, our main contributions are:
\begin{itemize}
    \item A methodology for synthesizing hardware-software leakage contracts from open-source processor designs.
    \item The definition of a concrete contract template for the {RISC-V} ISA and a corresponding test-case generation strategy.
    \item The implementation of a contract synthesis toolchain and its application to two open-source RISC-V processor designs.
\end{itemize}

\section{Preliminaries}

\subsection{Instruction Set Architectures}

\newcommand{\itSmallCaps}[1]{\textsl{\textsc{#1}}}
\newcommand{\isa}{\itSmallCaps{ISA}}
\newcommand{\arch}{\itSmallCaps{Arch}}

Instruction-set architectures (ISAs) define the software-visible interface of a processor.
For instance, they define the architectural state, the set of supported instructions, and how these instructions modify the architectural state.
We model an ISA as a state machine capturing the execution of programs one instruction at a time. 
Formally, an ISA is a function $\isa: \arch \rightarrow \arch$ that maps each architectural state $\sigma \in \arch$ to its successor $\isa(\sigma)$, obtained by executing the instruction to be executed in~$\sigma$.
To capture an entire execution, we denote by $\isa^*(\sigma)$ the sequence of  states reached from $\sigma$ by successive application of $\isa$, \ie, $\isa^*(\sigma) := [\sigma_0, \sigma_1, \sigma_2, \ldots]$ with $\sigma_0 = \sigma$ and $\sigma_{i+1} = \isa(\sigma_i)$ for all~$i \geq 0$.

\subsection{Microarchitectures}
\label{sec:muarch}

\newcommand{\muarch}{\itSmallCaps{Impl}}
\newcommand{\mustates}{\itSmallCaps{ImplState}}
\newcommand{\mumustates}{\mu\itSmallCaps{Arch}}
\newcommand{\attacker}{\mu\itSmallCaps{Atk}}
\newcommand{\atkobservations}{\itSmallCaps{AtkObs}}
\newcommand{\contract}{\itSmallCaps{Ctr}}
\newcommand{\ctrobservations}{\itSmallCaps{CtrObs}}

Microarchitectures are concrete implementations of an ISA and often contain complex performance-enhancing optimizations. 
They operate at cycle granularity rather than at instruction granularity. 
Thus, we model them as state machines that define how the processor's state evolves cycle-by-cycle.
Formally, a microarchitecture is a function $\muarch: \mustates \rightarrow \mustates$ on the set of microarchitectural states~$\mustates = \arch \times \mumustates$ that captures how the microarchitectural state evolves from one cycle to the next.
Each microarchitectural state  $\sigma \in \mustates$ consists of an architectural part $\sigma_\isa \in \arch$ and a microarchitectural part  $\sigma_\muarch \in \mumustates$ modeling the state of microarchitectural components like caches and predictors. 
Similarly to $\isa^*(\sigma)$, $\muarch^*(\sigma)$ is the sequence of microarchitectural states reached from $\sigma$ by successive applications of $\muarch$.\looseness=-1 

\subsection{Microarchitectural Attackers}

Programs that operate on secrets may leave traces of these secrets in the microarchitectural state.
Microarchitectural attacks extract information about secrets from the microarchitectural state by leveraging software-visible side-effects~\cite{spectre2019,meltdown2018,Bulck2018,Yarom14,RIDL}, mostly affecting a program's execution time.
In this paper, we consider (passive) microarchitectural attackers that can  extract information from the microarchitectural state.
Formally, we model a microarchitectural attacker as a function $\attacker: \mustates \rightarrow \atkobservations$ that maps microarchitectural states to attacker observations. 
Common attacker models, like the one exposing the timing of instruction retirement~\cite{Tsunoo03} or the one exposing the final state of caches~\cite{Yarom14,Doychev2015}, can be instantiated in our setting.
Given an attacker model, 
two executions $\muarch^*(\sigma)$ and $\muarch^*(\sigma')$ are \emph{attacker distinguishable} if $\attacker(\muarch^*(\sigma)) \neq \attacker(\muarch^*(\sigma'))$, where we lift $\attacker$ to sequences by applying it to each element of the sequence.

\subsection{Hardware-Software Leakage Contracts}

The goal of leakage contracts~\cite{contracts2021} is to capture microarchitectural side-channel leakage at the ISA level to allow reasoning about side-channel security of programs without having to explicitly consider their microarchitectural execution.
To this end, contracts associate  ISA-level executions with leakage traces, \ie, sequences of leakage observations.

A \emph{contract} is a function $\contract: \arch \rightarrow \ctrobservations$ mapping architectural states to leakage observations.
As an example, a contract can expose the addresses of memory instructions to capture leaks via the data cache.
Two architectural states $\sigma,\sigma'$ are \emph{contract distinguishable} if $\contract(\isa^*(\sigma)) \neq \contract(\isa^*(\sigma'))$, where we lift $\contract$ to sequences by applying it to all elements. 
Microarchitecture~$\muarch$ \emph{satisfies} contract~$\contract$ for instruction-set architecture~$\isa$ under attacker model $\attacker$ if all contract-indistinguishable executions are also attacker indistinguishable: 
\begin{multline*}
    \forall \sigma, \sigma' \in \mustates: \sigma_\muarch = \sigma'_\muarch \implies \\
        \contract(\isa^*(\sigma_\isa)) = \contract(\isa^*(\sigma'_\isa)) \implies \\\attacker(\muarch^*(\sigma)) = \attacker(\muarch^*(\sigma'))
\end{multline*}
Note that we require the microarchitectural parts of $\sigma$ and $\sigma'$ to initially be the same.
Otherwise, the executions could be attacker distinguishable due to initial differences rather than due to leakage during the execution.

The benefit of contract satisfaction is that it allows reasoning about side-channel security directly at the ISA level:
A program that does not leak any secret information according to the contract is also guaranteed not to leak any secret information to attackers on \emph{any} microarchitecture that satisfies the contract~\cite{contracts2021}.\looseness=-1

\section{Contract-Synthesis Methodology}

\newcommand{\precision}{\textit{Precision}\xspace}
\newcommand{\sensitivity}{\textit{Sensitivity}\xspace}
\newcommand{\indist}{\textit{Indist}\xspace}
\newcommand{\dist}{\textit{Dist}\xspace}
\newcommand{\distinguishing}{\textit{distinguishing}\xspace}
\newcommand{\template}{\textit{T}\xspace}
\newcommand{\testcases}{\textit{TC}\xspace}

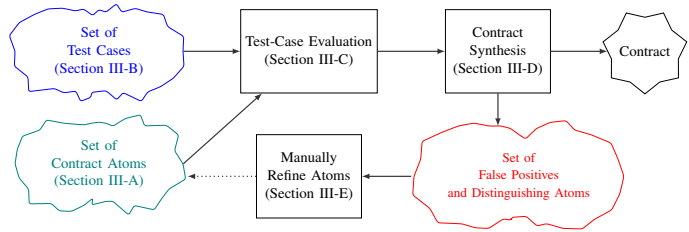
\begin{figure}[t]
    \centering\resizebox{\columnwidth}{!}{\begin{tikzpicture}[every text node part/.style={align=center}]

\tikzstyle{arrow} = [draw, thick, color=black!80, font=\footnotesize\sffamily]

\set{-1}{0}{1cm}{2cm}{2mm}{blue}{TC}{Set of\\ Test Cases\\(Section \ref{sec:testcasegeneration})};
\set{-1}{-2.7}{1cm}{2cm}{2mm}{teal}{ATOMS}{Set of\\ Contract Atoms\\(Section~\ref{sec:contracttemplate})};

\node[draw,rectangle,minimum width=25mm, minimum height=20mm] (ANALYZE) at (4,0) {Test-Case Evaluation\\(Section~\ref{sec:testcaseevaluation})};
\node[draw,rectangle,minimum width=25mm, minimum height=20mm] (SYNTHESIS) at (8.5,0) {Contract\\Synthesis\\(Section~\ref{sec:contractsynthesis})};
\node[draw,star,star points=7,star point ratio=0.8] (CONTRACT) at (12,0) {Contract};
\set{9}{-3}{1cm}{2.5cm}{1.5mm}{red}{FP}{\small Set of\\\small False Positives\\ \small and Distinguishing Atoms};
\node[draw,rectangle,minimum width=25mm, minimum height=20mm] (REFINE) at (4,-3) {Manually\\Refine Atoms\\(Section~\ref{sec:templaterefinement})};

\draw[->,arrow] (TC.east) -- (ANALYZE);
\draw[->,arrow] (ATOMS.east) -- (ANALYZE);

\draw[->,arrow] (ANALYZE.east) -- (SYNTHESIS.west);
\draw[->,arrow] (SYNTHESIS.east) --(CONTRACT);

\draw[->,arrow] (SYNTHESIS) -- ($(-.5,0.2) +(FP.north)$);
\draw[->,arrow] (FP.west) -- (REFINE.east);
\draw[->, arrow, dotted] (REFINE.west) -- ($(.1,-.3) +(ATOMS.east)$);

\end{tikzpicture}}
    \vspace{-7mm}
    \caption{High-level steps of our contract-synthesis methodology.}
    \vspace{-4mm}
    \label{fig:high-level}
\end{figure}
Our goal is to synthesize contracts directly from a processor's register-transfer level (RTL) design.
Figure~\ref{fig:high-level} overviews our contract-synthesis methodology, whose steps we describe next.

\subsection{Contract Templates}\label{sec:contracttemplate}
The basic building blocks of contracts are \emph{contract atoms}.
Contract atoms capture potential leakage at the instruction level.
Formally, a contract atom~$A$ is a tuple $(\pi_A, \tau_A, \phi_A)$ where $\pi_A : \arch \rightarrow \mathbb{B}$ determines whether the contract observation is applicable in a particular architectural state, $\tau_A$ is an identifier for the leakage source, and $\phi_A : \arch \rightarrow O_A$ is the observation function that maps an architectural state to the atom's observation.
For example, a contract atom $A$ could expose the divisor of division instructions:
$\pi_A$ would hold whenever the instruction to execute is a division, $\tau_A$ would identify the leakage source, 
and $\phi_A$ would return the divisor's value.\looseness=-1

A \emph{contract template}~$\template$ is a set of contract atoms.
A subset $S \subseteq T$ of the template defines a candidate \emph{contract} $\contract_S$ as follows:
$\contract_S(\sigma) = \{(\tau_A, \phi_A(\sigma)) \mid A \in S \wedge \pi_A(\sigma)\}$, \ie, it exposes the leakage source and the observation for each atom that is applicable in a given state.
Note that multiple atoms may generate observations for the same leakage source.

\subsection{Test-Case Generation}\label{sec:testcasegeneration}

Our goal is to find a contract that is satisfied by the microarchitecture while being as precise as possible, \ie, distinguishing as few executions as possible.
This requires (1)~to determine which contracts from the template are satisfied by the microarchitecture, and (2) to measure their precision.
Ideally, one would like to formally verify contract satisfaction.
However, contract verification techniques~\cite{wang2023specification} do not yet scale to processors of the complexity of, \eg, the CVA6 core.
Thus, we resort to a testing-based approach and evaluate contract satisfaction on systematically-generated test cases.
Formally, a \emph{test case} is a pair $(\sigma, \sigma')$ of microarchitectural states such that $\sigma_\muarch = \sigma'_\muarch$. 
A test case is \emph{attacker distinguishable} if $\attacker(\muarch^*(\sigma)) \neq \attacker(\muarch^*(\sigma'))$.
A test case is \emph{atom distinguishable} for an atom $A$ if $\contract_{\{A\}}(\isa^*(\sigma_\isa)) \neq \contract_{\{A\}}(\isa^*(\sigma'_\isa))$.
Note that a test case is contract distinguishable for contract $\contract_S$ if it is atom distinguishable for at least one atom in~$S$.
Thus, checking atom distinguishability for all atoms is sufficient to characterize contract distinguishability for the entire contract template.\looseness=-1

Test cases also allow to measure the precision of a contract.
For this, we adopt the standard notion of precision used in the evaluation of binary classifiers:
$\textit{Precision} = \frac{\mathit{TP}}{\mathit{TP}+\mathit{FP}}$,
where~$\mathit{TP}$ is the number of true positives, \ie, test cases that are contract distinguishable \emph{and} attacker distinguishable, and $\mathit{FP}$ is the number of false positives, \ie, test cases that are contract distinguishable \emph{but not} attacker distinguishable. 
Higher precision contracts are desirable as they rule out fewer programs at the contract level that could actually be executed securely on the processor.\looseness=-1
Our methodology requires a human expert to devise a test-case generation strategy used to generate the set of test cases~\testcases. 

\subsection{Test-Case Evaluation}\label{sec:testcaseevaluation}
Given a set of test cases $\testcases$, we need to determine for each test case (1) whether it is attacker distinguishable, and (2) which contract atoms distinguish it.
Attacker distinguishability can be determined via simulation 
of the microarchitecture with a suitable attacker model.
Similarly, atom distinguishability can be determined by evaluating all contract atoms in parallel on top of a simulation of the instruction set architecture.

The test-case evaluation phase has two outputs:
\begin{inparaenum}[(1)]
    \item The set of attacker-distinguishable test cases $\dist \subseteq \testcases$.
    \item For each test case $t$, the set of distinguishing atoms $\distinguishing(t) \subseteq T$.
\end{inparaenum}

\subsection{Contract Synthesis}\label{sec:contractsynthesis}

We now show how to use integer linear programming (ILP) to synthesize a contract from the template that distinguishes all attacker-distinguishable test cases \testcases and maximizes precision.\looseness=-1

The ILP uses a boolean variable $s_A$ for each atom $A$ in the contract template $\template$ to encode whether the atom is selected to be part of the synthesized contract or not.
Further, the ILP uses a boolean variable $c_t$ for each attacker-indistinguishable test case $t \in \indist = \testcases \setminus \dist$, which, using constraints detailed below, is forced to be $1$ for test cases that are contract distinguishable and thus are false positives.

Maximizing precision is equivalent to minimizing the number of false positives, as the number of true positives is the same for all correct contracts.
Thus, the objective function of the ILP is $\min \sum_{t \in \indist} c_t$.
To ensure that only correct contracts are considered, we introduce the following constraint for each attacker-distinguishable test case $t \in \dist$:
$\sum_{A \in \distinguishing(t)} s_A \geq 1$,
\ie, at least one atom that distinguishes $t$ must be selected for the contract.
For each attacker-indistinguishable test case $t \in \indist$ and each contract atom $A \in \distinguishing(t)$ we further introduce the constraint:
$s_A \leq c_t$.
This ensures that $c_t$ can only be $0$ for test cases that are not contract distinguishable.

From the ILP's solution, we extract the synthesized contract via the variables $s_A$ and the false-positive test cases via the variables $c_t$.\looseness=-1

\subsection{Refinement of the Contract Template}\label{sec:templaterefinement}

In addition to returning a contract that maximizes precision, our implementation also returns a ranking of the contract atoms according to the number of false positives caused by their inclusion in the contract and the corresponding test cases.
Inspecting these test cases allows a human expert to identify contract atoms that should be refined to obtain a more precise contract.\looseness=-1

\section{Instantiating the Methodology for RISC-V}

We instantiated the above methodology for the RISC-V instruction set, more precisely its \texttt{I} and \texttt{M} subsets. 

\subsection{RISC-V Contract Template}
\label{sec:riscv-template}
Instructions of different type often show different leakage behavior, \eg, while divisions may leak the second operand, additions usually do not.
Thus, our contract template contains a contract atom for each instruction type and each applicable potential leakage source.
For each atom, $\pi_A$ determines whether the current instruction has the given type, $\tau_A$ is an identifier for the leakage source, and $\phi_A$ extracts the corresponding leakage from the architectural state.
For example, the contract atom $(\pi^\texttt{DIV}, \texttt{REG\_RS2}, \phi^\texttt{REG\_RS2})$, where $\pi^\texttt{DIV}$ detects divisions and $\phi^\texttt{REG\_RS2}$ determines the value of register \texttt{RS2}, allows to capture leakage of the second operand of division instructions.
Similarly, the contract atom $(\pi^\texttt{ADD}, \texttt{REG\_RS2}, \phi^\texttt{REG\_RS2})$ allows to capture leakage of the second operand of addition instructions, sharing the same leakage source as the corresponding division atom.
\looseness=-1

We first defined a base template capturing the architectural state that directly influences the execution of an instruction:
\begin{asparaitem}
    \item \textbf{Instruction leakages (\texttt{IL})} expose values from an instruction's encoding: The operation \texttt{OP}, the destination and source registers \texttt{RD}, \texttt{RS1}, and \texttt{RS2}, and the immediate value~\texttt{IMM}.
    \item \textbf{Register leakages (\texttt{RL})} expose the values of registers: The values of the source registers \texttt{REG\_RS1} and \texttt{REG\_RS2} before execution and the final value of the destination register \texttt{REG\_RD}.
    \item \textbf{Memory leakages (\texttt{ML})} expose the memory addresses and memory contents accessed by an instruction:
    \texttt{MEM\_R\_ADDR} and \texttt{MEM\_W\_ADDR} expose the accessed addresses, whereas \texttt{MEM\_R\_DATA} and \texttt{MEM\_W\_DATA} expose the accessed content.
\end{asparaitem}
During evaluation, we noticed that, while the above categories are sufficient, the precision of synthesized contracts can be improved by adding the following contract atoms:
\begin{asparaitem}
    \item \textbf{Alignment leakages (\texttt{AL})} expose the alignment of a memory access: \texttt{IS\_WORD\_ALIGNED} exposes whether the last two bits of the memory address are $00$ and \texttt{IS\_HALF\_ALIGNED} exposes whether the last two bits of the address are not $11$. 
    \item \textbf{Branch leakages (\texttt{BL})} expose whether a branch is taken or not: This can be \texttt{BRANCH\_TAKEN} for taken branches and \texttt{NEW\_PC} to indicate the target of the branch. 
    The latter one also applies to unconditional jumps.
    \item \textbf{Data-dependency leakages (\texttt{DL})} expose data dependencies in the program: \texttt{RAW\_RS1\_n}, \texttt{RAW\_RS2\_n}, \texttt{RAW\_RD\_n},  and \texttt{WAW\_n}, respectively, indicate Read-After-Write dependencies on register \texttt{RS1}, \texttt{RS2}, and \texttt{RD} and Write-After-Write dependencies within a distance of $n$ instructions. 
\end{asparaitem}

We remark that not all atoms are applicable to every instruction type, \eg, some instructions do not have an immediate value. 
With a maximum distance of $n=4$, the above template results in a total of 762 atoms.

\subsection{Test-Case Generation}

The set of test cases used for the simulation affects the synthesized contract, since the synthesis algorithm cannot account for leakages that are not exposed by the test cases. 
Thus, ideally the test-case generation algorithm should consider potential microarchitectural implementations and their potential leakage.\looseness=-1

The test-case generation method we use is simple but has shown effective on the tested targets. Note that we only generate the architectural state and fix the initial microarchitectural part of the microarchitectural state to allow reusing test cases for different microarchitectures. 
Each test case consists of two programs which aim to be differentiable by one specific contract atom.\looseness=-1

Each program consists of three parts, of which only the second part differs among the two programs in the test case:
\begin{inparaenum}[(1)]
    \item First, we initialize every architectural register to a randomly selected value. 
    \item Next, we try to trigger the leakage of the contract atom we want to test for. 
    For this, we generate a random instance of the given instruction type. 
    We then derive two sequences of instructions that could be differentiated by the given atom, \eg, if we want to test whether the immediate value leaks, we alter the immediate value or if we want to test whether the memory contents that are read leak, we insert an earlier instruction in both programs writing different values to the same address.
    \item Finally, we append randomly selected instructions that aim to surface the leakage and to make sure all instructions from (2) are executed completely.
\end{inparaenum}

\subsection{Attacker Model}\label{sec:attackermodel}

We consider an attacker that can observe the timing of instruction retirements at cycle granularity.
Our implementation of this attacker model relies on the \mbox{RISC-V} Formal Interface (RVFI)~\cite{wolf2017riscv-formal}, a standard interface for \mbox{RISC-V} processors exposing information about the execution of programs, including the cycles at which instructions retire.
RVFI simplifies the interaction with the microarchitecture and allows reusing most of our implementation for any processor implementing it.

By instantiating two instances of the respective core and simulating the execution of both programs of a test case in parallel, our implementation determines attacker distinguishability by comparing the cycles at which instructions retire.
Other attacker models could be implemented either using the RVFI or by exposing the required signals directly from the design.

\subsection{Identifying Distinguishing Atoms}\label{sec:distinguishingatoms}

To determine the distinguishing atoms for a test case, we need to evaluate all contract atoms on top of an architectural simulation of both programs of a test case.

However, as long as the microarchitecture correctly implements the ISA, the sequence of architectural states can also be extracted from the microarchitectural states upon instruction retirement~\cite{wang2023specification}.
Thus, our implementation piggy backs on the microarchitectural simulation from \S\ref{sec:attackermodel} and it uses the RVFI to extract the architectural state whenever an instruction retires.
The simulation produces a VCD waveform, from which we derive a test case's distinguishing atoms. 
Adaptation to other cores that support the RVFI should be straightforward.

\section{Experimental Evaluation}
\newcommand{\cmark}{\ding{51}}
\newcommand{\xmark}{\ding{55}}
\newcommand{\ctrp}{$\circ$\xspace}
\newcommand{\ctrf}{\cmark\xspace}
\newcommand{\ctrn}{\xmark\xspace}
\renewcommand{\ctrp}{$\exists$\xspace}
\renewcommand{\ctrf}{$\forall$\xspace}
\renewcommand{\ctrn}{$\nexists$\xspace}
\def\downcirc{\raisebox{1pt}{\scalebox{0.6}{\rotatebox[origin=c]{90}{\LEFTcircle}}}}
\makeatletter
\DeclareRobustCommand{\circbullet}{\mathbin{\vphantom{\circ}\text{\circbullet@}}}
\newcommand{\circbullet@}{
  \check@mathfonts
  \m@th\ooalign{
    \clipbox{0 0 0 {\dimexpr\height-\fontdimen22\textfont2}}{$\bullet$}\cr
    $\circ$\cr
  }
}
\renewcommand{\ctrp}{{\Large$\circbullet$}\xspace}
\renewcommand{\ctrf}{{\Large$\bullet$}\xspace}
\renewcommand{\ctrn}{{\Large$\circ$}\xspace}
\newcommand{\ctra}{-}

We evaluate our methodology on two open-source \mbox{RISC-V} cores, Ibex~\cite{ibex} and CVA6~\cite{cva6}, both in a configuration implementing \texttt{RV32IMC}, for which we obtain precise contracts. 
Additionally, we analyze the impact of the test-case generation method and the granularity of the contract template.

\subsection{Experimental Setup}
To enable the simulation of test cases we first convert the processor sources from SystemVerilog
to standard Verilog 
    using sv2v~\cite{snow2019sv2v} for Ibex and Yosys~\cite{wolf2016Yosys} with the Yosys-systemverilog plugin~\cite{rakoczy2021systemverilog} for CVA6. 
and then compile and execute the two cores embedded in a testbench using Icarus Verilog~\cite{williams2000iverilog}. 
In both cases, some manual adjustments to the processors were needed to eliminate unsupported SystemVerilog constructs. 

Both cores support the RVFI, which we use to extract the attacker observations and distinguishing atoms as described in \S\ref{sec:attackermodel} and \S\ref{sec:distinguishingatoms}.
The contract synthesis algorithm is implemented in Java using Google's OR-Tools~\cite{google2010ortools} ILP solver.

\begin{figure}
    \centering
    \begin{tikzpicture}
    \begin{axis}[
            xmin=0,
            xmax=100000,
            ymin=0,
            ymax=1,
            xtick={0,25000,50000,75000,100000},
            scaled ticks=false,
            tick label style={/pgf/number format/fixed},
            legend style={yshift=-0.75cm,xshift=-2mm,nodes={scale=0.45,transform shape}},
            width=\columnwidth,
            height=4cm,
            tick label style={font=\tiny},
            label style={font=\tiny},
        ]
        \addlegendentry{IL + RL + ML}
        \addlegendentry{IL + RL + ML + AL}
        \addlegendentry{IL + RL + ML + AL + BL}
        \addlegendentry{IL + RL + ML + AL + BL + DL}
        \addplot[line width=1pt,solid,color=gray]
            table[x=index,
                y=precision-base,
                col sep=comma,
                each nth point=1,
                filter discard warning=false,
                unbounded coords=discard]
                {data/ibex.csv};
        \addplot[line width=1pt,solid,color=red]
            table[x=index,
                y=precision-base-align,
                col sep=comma,
                each nth point=1,
                filter discard warning=false,
                unbounded coords=discard]
                {data/ibex.csv};
        \addplot[line width=1pt,solid,color=orange]
            table[x=index,
                y=precision-base-align-branch,
                col sep=comma,
                each nth point=1,
                filter discard warning=false,
                unbounded coords=discard]
                {data/ibex.csv};
        \addplot[line width=1pt,solid,color=blue]
            table[x=index,
                y=precision-base-align-branch-deps,
                col sep=comma,
                each nth point=1,
                filter discard warning=false,
                unbounded coords=discard]
                {data/ibex.csv};
    \end{axis}
\end{tikzpicture}
    \vspace{-5mm}
    \caption{Precision of contracts (y-axis) w.r.t. 2,000,000 test cases for different contract templates starting from the base contract (\texttt{IL+RL+ML}) depending on the number of test case (x-axis) used for contract synthesis.}
    \label{fig:ibex-precision}
    \vspace{-2mm}
\end{figure}
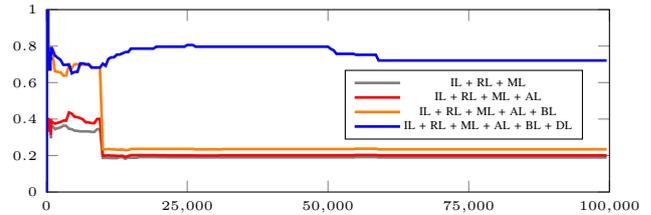

\begin{figure}
    \centering
    \begin{tikzpicture}
    \begin{axis}[
            xmin=0,
            xmax=100000,
            ymin=0,
            xtick={0,1,10,100,1000,10000,100000},
            ytick={0,0.25,0.5,0.75,1},
            scaled x ticks=false,
            tick label style={/pgf/number format/fixed},
            legend style={nodes={scale=0.5, transform shape}},
            width=\columnwidth,
            height=3.7cm,
            xmode=log,
            log ticks with fixed point,
            tick label style={font=\tiny},
            label style={font=\tiny},
        ]
        \addplot[line width=1pt,solid,color=blue]
            table[x=index,
                y=sensitivity,
                col sep=comma,
                ]
                {data/ibex.csv};
    \end{axis}
\end{tikzpicture}
    \vspace{-5mm}
    \caption{Sensitivity of contracts (y-axis) w.r.t. 2,000,000 test cases using the full contract template (\texttt{IL+RL+ML+AL+BL+DL}) depending on the number of test cases (x-axis) used for contract synthesis. 
    Note the logarithmic x-axis.
    }
    \vspace{-4mm}
    \label{fig:ibex-sensitivity}
\end{figure}
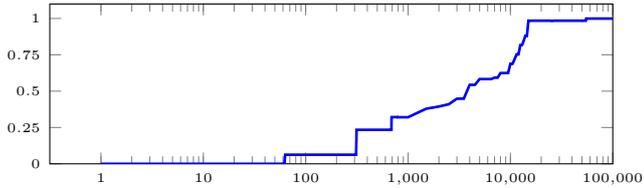

\subsection{Ibex Core}
We first synthesized a contract for the Ibex core using the base contract template described in Section~\ref{sec:riscv-template}. 
To this end, we used 100,000 test cases for synthesis and evaluated the performance of the obtained contract using an independent set of 2,000,000 test cases. 
The results allowed us to refine the contract template (as described in Section~\ref{sec:riscv-template}) by adding alignment leakages (\texttt{AL}), branch leakages (\texttt{BL}), and data-dependency leakages (\texttt{DL}).

Figure~\ref{fig:ibex-precision} shows the precision of the obtained contracts for different contract templates in terms of the number of test cases used for synthesis (from 0 to 100,000).
Some leakage sources are only discovered after a while, \eg, the first data-dependency leakages are discovered after about 10,000 test cases, which explains the drop in precision for templates that do not include \texttt{DL} at that point.
We observe that the refined templates lead to an increase in precision as they allow for more fine-grained leakage contracts. 
Data-dependency leakages (\texttt{DL}), in particular, enable significant precision improvements.

Figure~\ref{fig:ibex-sensitivity} shows the sensitivity of the synthesized contracts for the full contract template depending on the number of test cases used for synthesis. 
We adopt the standard notion of sensitivity used in the evaluation of binary classifiers:
$\textit{Sensitivity} = \frac{\mathit{TP}}{\mathit{TP}+\mathit{FN}}$, where $\mathit{FN}$ is the number of false negatives.
Initially, sensitivity increases rapidly, as additional test cases frequently reveal new sources of leakage. 
After the first 15,000 test cases, however, the curve flattens.
The final contract has a sensitivity of 99.93\%. 

Finally, we synthesized a contract, summarized in Table~\ref{table:ibex},  using the larger set of 2,000,000 test cases.
In total, the synthesized contract includes 82 atoms. 
\ctrf indicates that all instructions in this category show leakages in the respective category (as introduced in Section~\ref{sec:riscv-template}), 
\ctrn indicates that no leakages were found, 
\ctrp indicates that there are some leakages in this category, but not all instructions have these leakages, and
\ctra{} indicates that the atom does not apply to this category.
The distance $n$ in the \texttt{DL} category is always 1.

Next, we overview some of our findings:
(1) Our experiments and the synthesized contracts show that the Ibex core leaks whether memory accesses are aligned or not.
Indeed, the documentation of the Ibex core confirms that requests on the memory interface are always aligned to a word boundary, which is compatible with our observation, but the impact on information leakage had not been established before.
(2) We also discovered that the timing of branch instructions depend on whether the branch is taken or not taken even if the branch target is the same in both cases, \eg, \texttt{BEQ r1 r2 4} jumps to the next instruction independently of \texttt{r1} and \texttt{r2}.

\begin{table}
    \centering
    \vspace{-5mm}
    \caption{Synthesized contract for the Ibex processor.
    }
    \vspace{-1mm}
    \centering
    \begin{tabular}{lccccccc}
        \toprule
                                & IL        & RL        & ML        & AL        & BL        & DL        \\
        \midrule
        Arithmetic instructions & \ctrp{}   & \ctrp{}   & \ctra{}   & \ctra{}   & \ctra{}   & \ctrp{}   \\
        Division, Remainder     & \ctrn{}   & \ctrp{}   & \ctra{}   & \ctra{}   & \ctra{}   & \ctrp{}   \\
        Multiplication          & \ctrp{}   & \ctrn{}   & \ctra{}   & \ctra{}   & \ctra{}   & \ctrf{}   \\
        \midrule 
        Loads                   & \ctrp{}   & \ctrn{}   & \ctrn{}   & \ctrf{}   & \ctra{}   & \ctrn{}   \\
        Stores                  & \ctrp{}   & \ctrn{}   & \ctrn{}   & \ctrn{}   & \ctra{}   & \ctrn{}   \\
        \midrule
        Branch instructions     & \ctrp{}   & \ctrn{}   & \ctra{}   & \ctra{}   & \ctrf{}   & \ctrn{}   \\
        \bottomrule
    \end{tabular}
    \label{table:ibex}
\end{table}

\subsection{CVA6 Core}
We similarly analyzed the CVA6 core. 
A summary of the contract obtained using 500,000 test cases is shown in Table~\ref{table:cva6}. 
In total the synthesized contract consists of 77 atoms.

The CVA6 core uses a more complex memory interface that does not expose anything about a specific memory access in the analyzed setup.
We remark that, even though the CVA6 features a simple branch predictor, the contract template originally composed for the Ibex, was sufficient to capture the CVA6's leakage.\looseness=-1

As the CVA6 core has a deeper and more complex pipeline than the Ibex,
data and control dependencies can have a more pronounced effect in CVA6, which is reflected in the synthesized contract:
While the data dependencies all have a distance of $n=1$ as forwarding is effective here, we observe distances of up to $n=4$ due to control dependencies upon branch instructions. 

\begin{table}
    \vspace{-3mm}
    \centering
    \caption{Synthesized contract for the CVA6 processor.
    }
    \vspace{-1mm}
    \centering
    \begin{tabular}{lccccccc}
        \toprule
                                & IL        & RL        & ML        & AL        & BL        & DL        \\
        \midrule
        Arithmetic instructions & \ctrp{}   & \ctrp{}   & \ctra{}   & \ctra{}   & \ctra{}   & \ctrp{}   \\
        Division, Remainder     & \ctrp{}   & \ctrp{}   & \ctra{}   & \ctra{}   & \ctra{}   & \ctrp{}   \\
        Multiplication          & \ctrn{}   & \ctrp{}   & \ctra{}   & \ctra{}   & \ctra{}   & \ctrp{}   \\
        \midrule
        Loads                   & \ctrp{}   & \ctrn{}   & \ctrn{}   & \ctrn{}   & \ctra{}   & \ctrp{}   \\
        Stores                  & \ctrn{}   & \ctrp{}   & \ctrn{}   & \ctrn{}   & \ctra{}   & \ctrn{}   \\
        \midrule
        Branch instructions     & \ctrn{}   & \ctrn{}   & \ctra{}   & \ctra{}   & \ctrf{}   & \ctrp{}   \\
        \bottomrule
    \end{tabular}
    \label{table:cva6}
\end{table}

\subsection{Computation Time}
Table~\ref{table:stats} compares the runtime of the contract synthesis algorithm for the two cores. The test-case simulation for CVA6 is much slower than on Ibex due to the higher complexity of the design and due to the translation into standard Verilog with Yosys. 
However, the contract synthesis algorithm is still able to synthesize a contract for  CVA6 in a reasonable amount of time.\looseness=-1

\begin{table}
    \centering
    \vspace{-5mm}
    \caption{Performance measurements for contract synthesis using 100,000 test cases on an AMD Ryzen Threadripper PRO 5995WX with \SI{512}{\giga\byte} of RAM using up to 128 threads.
    }
    \vspace{-2mm}
    \begin{tabular}{lrr}
        \toprule
                                            & Ibex  & CVA6  \\
        \midrule
        Compilation of the testbench        & \SI{7.2}{\second} & \SI{2123}{\second} \\
        Simulation of a single test case    & \SI{0.2}{\second} & \SI{88}{\second}   \\
        Extraction of distinguishing atoms  & \SI{175}{\milli\second} & \SI{75}{\milli\second}  \\
        Computation of the contract         & \SI{15.6}{\second} & \SI{16}{\second}   \\
        \midrule
        Overall computation time            & \SI{5.3}{\minute}& \SI{1175}{\minute} \\
        \bottomrule
    \end{tabular}
    \label{table:stats}
    \vspace{-2mm}
\end{table}

\section{Related Work}

\newcommand{\compactparagraph}[1]{\smallskip\noindent\textbf{#1:}}

\compactparagraph{Leakage contracts}
Leakage contracts~\cite{contracts2021}, which our methodology builds on, are an abstraction for capturing timing leaks at ISA level.
The leaks captured in a contract are formally connected with the actual timing leaks in a hardware implementation.
There are several tools for reasoning about contracts:
for verifying contract satisfaction against RTL processor designs~\cite{wang2023specification,powercontracts}, for detecting contract violations for black-box CPUs~\cite{oleksenko2022revizor,oleksenko2023hide,Nemati2020a,buiras2021micro,Hofmann23}, and for verifying program security w.r.t. a given contract~\cite{spectector2020,fabian202automatic}.
Contracts synthesized using our methodology may serve as inputs to these tools.

Beyond leakage contracts, 
several proposals extend ISAs with leakage guarantees.
Yu et al.~\cite{Yu19} propose a data-oblivious ISA extension that allows to specify whether or not instruction operands are safe (i.e., never leaked microarchitecturally). 
Similarly, CPU vendors implement data-oblivious execution modes that, when activated, ensure that the execution time of specific instructions is independent of their operands~\cite{inteldoit,armdit,riscvzkt}.
All these guarantees can be formalized as leakage contracts.\looseness=-1

\compactparagraph{Modeling timing leaks}
There are multiple formal approaches for studying timing leaks.
Most of them capture leaks at program level: from simple models associated with ``constant-time programming''~\cite{almeida2016verifying,molnar2005program} to more complex ones capturing leaks of transient instructions~\cite{spectector2020,patrignani2021exorcising,fabian202automatic,pitchfork, blade}.
Other approaches, instead, capture leaks on simplified processor models~\cite{pandorasbox,GuancialeBD20}. 
Our work enables synthesizing program-level models from actual processor implementations, while previous approaches rely on manually written models that have no formal connection to particular concrete processors.

\compactparagraph{Detecting leaks through testing}
Revizor~\cite{oleksenko2022revizor,oleksenko2023hide,Hofmann23} and Scam-V~\cite{Nemati2020a,buiras2021micro} search for contract violations 
for black-box CPUs and can only be applied post-silicon.
Other approaches~\cite{Weber2021,Gras2020,Moghimi2020a} detect leaks by analyzing hardware measurements without relying on leakage contracts but, again, apply only post-silicon.
Finally, SpecDoctor~\cite{hur2022specdoctor} and SIGFuzz~\cite{rajapakshasigfuzz} focus on detecting microarchitectural leaks in RTL designs in the pre-silicon phase.
In contrast to our work, they do not aim to characterize leakage at ISA level.
\vspace{-0.5mm}

\section{Conclusion}

ISA-level models of the information leaked microarchitecturally by processors are a critical component for the  development of systems resistant to microarchitectural attacks.
In this paper, we showed how to semi-automatically synthesize such models, in the form of hardware-software leakage contracts, directly from RTL processor designs.
This allowed us to derive ISA-level descriptions of leakage for two open-source RISC-V cores, Ibex and CVA6, which so far lacked a formal specification of their microarchitectural leakage properties.
The contract synthesis toolchain is available open source at \url{https://github.com/hw-sw-contracts/riscv-contract-synthesis} and archived at~\url{https://doi.org/10.5281/zenodo.10491534}.

\bibliographystyle{IEEEtranS}
\bibliography{references}

\end{document}